\documentclass[aps,prb,twocolumn,longbibliography,amsmath,amssymb,nofootinbib,longbibliography,10pt]{revtex4-2}
\usepackage{amsfonts,amssymb,amsmath}
\usepackage{graphics}
\usepackage{graphicx}
\usepackage{nicefrac}
\usepackage{cases}
\usepackage{mathrsfs}
\usepackage{enumerate}
\usepackage{mathtools}
\usepackage{textcomp}
\usepackage{verbatim}
\usepackage{bm}          
\usepackage{soul}
\usepackage{cancel}
\usepackage{upgreek}
\usepackage{txfonts}
\usepackage{color}
\usepackage[colorlinks=true, urlcolor=blue, linkcolor=blue]{hyperref}
\usepackage[papersize={8.5in,11in}]{geometry}
\usepackage{esint}
\newcommand{\be}{\begin{equation}}
\newcommand{\ee}{\end{equation}}
\usepackage{makecell}

\begin{document}

\title{Upper bound on the window of density occupied by microemulsion phases in 2D electron systems}
\author{Sandeep Joy}
\author{Brian Skinner}
\affiliation{Department of Physics, Ohio State University, Columbus, OH 43210, USA}
\date{\today}

\begin{abstract}

In two-dimensional electronic systems, direct first-order phase transitions are prohibited as a consequence of the long-range Coulomb interaction, which implies a stiff energetic penalty for macroscopic phase separation. A prominent proposal is that any direct first-order transition is instead replaced by a sequence of ``microemulsion" phases, in which the two phases are mixed in patterns of mesoscopic domains. In this note, we comment on the range $\Delta n$ of average electron density that such microemulsion phases may occupy.  We point out that, even without knowing the value of a phenomenological parameter associated with surface tension between the two phases, one can place a fairly strong upper bound on the value of $\Delta n$. We make numerical estimates for $\Delta n$ in the case of the Fermi liquid to Wigner crystal transition and find $\Delta n$ to be on the order of $10^7$\,cm$^{-2}$. This value is much smaller than the width of the phase transition observed in experiments, suggesting that disorder is a more likely explanation for the apparent broadening of the transition.
  
\end{abstract}
\maketitle

Over two decades ago, it was established that there can be no direct first-order transition between solid and liquid phases of a two-dimensional electron system (2DES) \cite{spivak2003phase, spivak2004phases, jamei2005universal}. The authors of Refs.~\cite{spivak2003phase, spivak2004phases, jamei2005universal} proposed that the nominal first-order transition is instead replaced by a series of ``microemulsion" phases, in which the two distinct states are blended together in spatially-structured mesoscopic domains (similar ideas appeared earlier in Ref.~\cite{Kivelson_quantization_1986}). Subsequent work has developed this idea further and explored its experimental implications \cite{ drag2005spivak, spivak2006transport, spivak2010colloqium}. The possibility of microemulsion phases is now commonly invoked when experiments show a phase transition in a 2DES that occupies a finite window of average electron density rather than an abrupt transition at a single density (e.g., Refs.~\cite{hatke2015microwave, zondiner2020cascade, rozenentropic2021}). In this short note, we comment on how large, exactly, can be the window of average density, $\Delta n$, occupied by microemulsion phases. After deriving a general bound on $\Delta n$, we focus specifically on the quantum melting transition between the Wigner crystal and Fermi liquid states.

\begin{figure}[htb]
\centering
\includegraphics[scale=0.3]{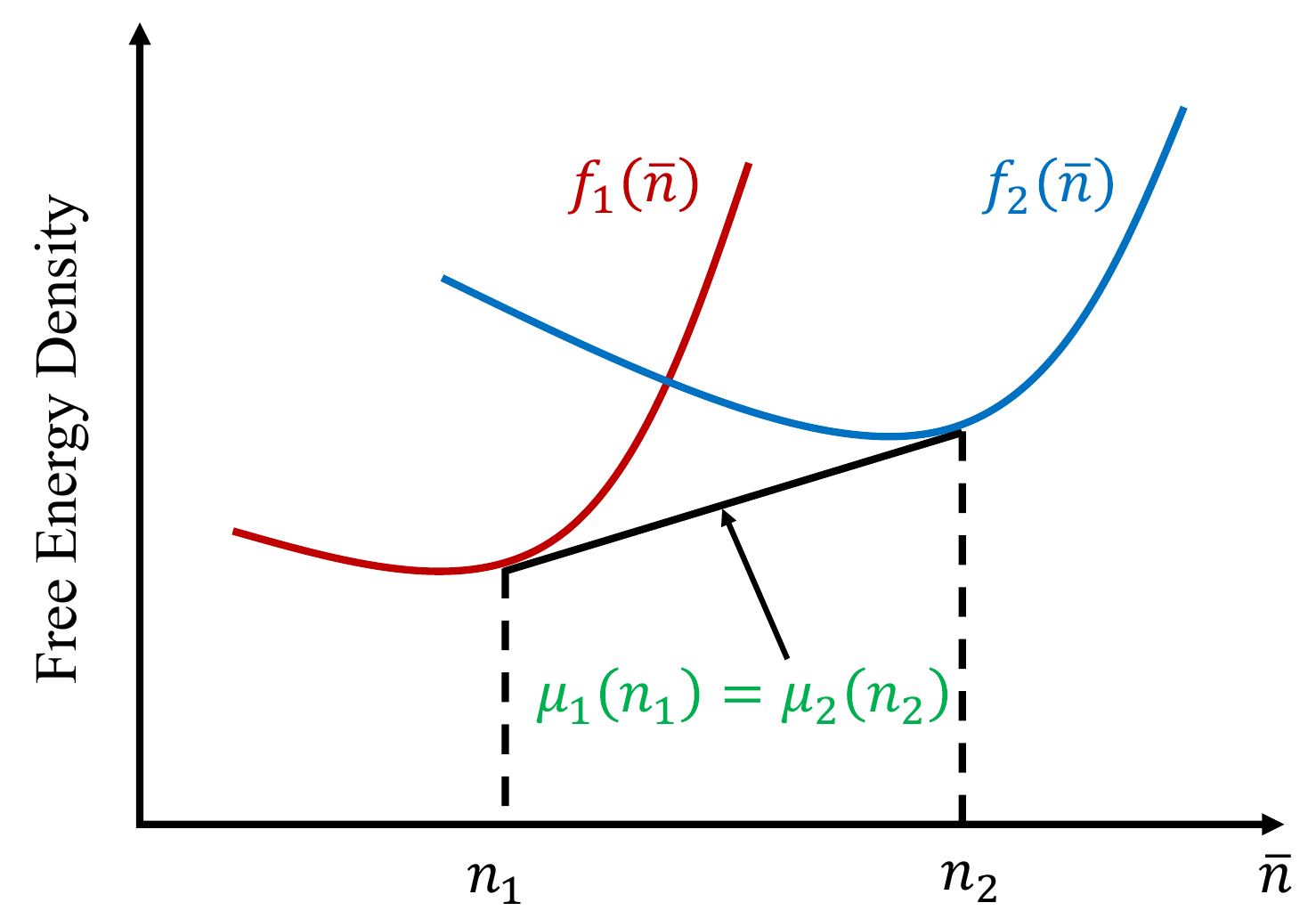}
\caption{Schematic depiction of the conventional Maxwell construction for a first-order phase transition. The red and blue curves represent the free energy per unit area of phases $1$ and $2$, respectively, as a function of the average density $\bar{n}$. The common tangent represents the coexistence line, in which there is a (macroscopic) phase separation. During this phase separation regime, both phases have the same chemical potential (the slope of the tangent).}
\label{fig:MC}
\end{figure}

A first-order phase transition and its associated (macroscopic) phase coexistence is typically understood using the framework of the Maxwell construction. Before considering the role of long-range Coulomb interactions between phase-segregated domains, we first briefly summarize the  conventional Maxwell construction. Consider a (first-order) phase transition between two different phases, which we label as 1 and 2. If $\varepsilon_{i}(n)$ denotes the energy per particle as a function of the particle density $n$ in phase $i = 1,2$, then at zero temperature, the free energy per unit area in phase $i$ is $f_i(n) = n \varepsilon_i(n)$.  In the usual Maxwell construction for phase coexistence, one considers that the total free energy for a fixed number of particles is minimized with respect to the area fraction $x$ that is occupied by one of the two phases (say, phase 2). That is, if $n_1, n_2$ denote the particle densities in spatial regions occupied by phases $1$ and $2$, respectively, then the region of average density which produces phase coexistence is found by minimizing the total free energy $(1-x)f_1(n_1) + x f_2(n_2)$ with respect to $x$ and $\delta n = n_2 - n_1$ for a fixed average density $\bar{n} = (1-x)n_1 + x n_2$. From this procedure, one can easily show that, within the regime of phase coexistence, the two phases have the same chemical potential $\mu_i = df_i(n) /dn$ (see Fig.~\ref{fig:MC}).~\footnote{The slope of the common tangent is given by $(f_2(n_2) - f_1(n_1) )/(n_2 - n_1) = \mu_2(n_2) = \mu_1(n_1)$. This condition is equivalent to the more widely recognized version of the Maxwell construction, known as the equal area rule: $\int_{n_{1}}^{n_{2}}(\mu(n) - \mu^\star)\, dn = 0$, where $\mu^\star$ is the chemical potential of the system during phase coexistence and $\mu(n)$ denotes the chemical potential of whichever uniform phase has lower energy at density $n$.}

In electron systems, however, there is an additional Coulombic energy cost associated with the difference between the local electron density ($n_1$ or $n_2$) and the charge density $+e\bar{n}$ of the positive background. (Throughout this note, we assume that the gate electrode containing the neutralizing background charge is much further away from the plane of the 2DES than the size of microemulsion domains; this assumption allows us to treat the neutralizing charge as spatially uniform. We comment at the end of this note about the effect of allowing the distance to the gate to be finite.) This Coulombic energy suppresses phase coexistence and generally prohibits the formation of macroscopic phase domains (since a region of size $L$ and charge density $\eta$ has a Coulombic self-energy per unit area that grows extensively with the region size, $\sim \eta^2 L$) \cite{fogler2001stripe}. Instead, one ends up with the microemulsion phase, where stripes or droplets of the less abundant phase are interspersed periodically among the more abundant phase. 

We emphasize that, while we discuss our results throughout this paper in terms of electrons, the physics we invoke is agnostic to whether the particles are bosons or fermions, or quantum or classical in nature. Our considerations apply generically to any 2D system with long-range Coulomb interactions.

\smallskip

\textit{\textbf{Argument for the absence of a direct first-order transition.}}
Before discussing an upper limit on $\Delta n$, we first briefly recapitulate the argument that there can be no direct first-order phase transitions in a 2DES (see Ref.~\cite{spivak2006transport} for a more complete review). The argument proceeds by considering the system at the critical density $n = n_c$ for which the two phases have equal energy (the nominal location of the first-order transition) and showing that one can construct a trial microemulsion state with the same average density but lower free energy than either pure phase. The existence of such a lower-energy state implies that the system must not pass directly from a pure phase 1 to a pure phase 2 as a function of increasing average density, and a direct first-order transition between the two pure phases is precluded. The argument is simplest if one assumes that the interaction between electrons is of the form $V(r)=k/r^\alpha$, where $1<\alpha<2$. The limit $\alpha \rightarrow 1$ is discussed below. 

Consider a trial state consisting of alternating stripes of phases 1 and 2, each with the same width $\ell$ (assumed to be much longer than the inter-electron spacing) and having uniform electron density $n_1 = n_c - \delta n$ (phase 1) and $n_2 = n_c + \delta n$ (phase 2), so that the global average density is $n_c$. The free energy per area of this phase, relative to either uniform phase at $n=n_c$, can be written
\be
\delta f = f_{\text{bulk}} + f_{\text{Coulomb}} + f_{\text{surface}}.
\ee

The first contribution, $f_{\text{bulk}}$, represents the change in free energy (per unit area) associated with shifting some electrons from phase 1 to phase 2, which has a lower chemical potential. To leading order in $\delta n$, $f_{\text{bulk}}$ is given by:
\be
f_{\text{bulk}} = -\frac{\delta n}{2} (\mu_1(n_c) - \mu_2(n_c)).
\ee

The second contribution, $f_{\text{Coulomb}}$, corresponds to the electrostatic energy cost associated with the two stripe regions having an overall net charge $\pm e \delta n$ per unit area:
\be
f_{\text{Coulomb}} = A(e\delta n)^2V(\ell)\ell^2,
\ee
where $A$ is a numerical constant of order unity. 

The final contribution, $f_{\text{surface}}$, represents the energy required to create an interface between the two phases:
\be
f_{\text{surface}} = \frac{\sigma}{\ell},
\ee
where $\sigma$ is the interfacial tension.

Keeping $\ell$ fixed, we minimize $\delta f$ with respect to $\delta n$, which gives
\be
\delta n = \frac{\mu_1(n_c) - \mu_2(n_c)}{4Ake^2\ell^{2-\alpha}}.
\ee
The corresponding (minimum) free energy change is given by:
\be
\delta f = -\frac{\left(\mu_1(n_c) - \mu_2(n_c)\right)^2}{16Ake^2\ell^{2-\alpha}} + \frac{\sigma}{\ell}.
\label{eq: deltafSK}
\ee
Notice that so long as $1<\alpha<2$, the change in free energy $\delta f$ is negative whenever $\ell$ is sufficiently long. Thus, we have constructed a trial microemulsion state that has a lower energy than either uniform state, and there must not be a direct first-order transition between two uniform states.

In the limit $\alpha \rightarrow 1$, both terms in Eq.~\ref{eq: deltafSK} are proportional to $1/\ell$, and the analysis is inconclusive. However, a more carefully constructed trial state, in which the electron density is allowed to vary within each stripe, still yields a negative value of $\delta f$ in the limit of very long $\ell$ \cite{jamei2005universal}. 

\smallskip

\textit{\textbf{Coulomb-frustrated phase coexistence.}} Because of the long-ranged Coulomb interactions between regions of different charge densities, phase coexistence cannot be understood without simultaneously considering the spatial structure of domains. This problem has been studied extensively for both 2D and 3D electron gases (3D in \cite{lorenzana2001phase1, lorenzana2001phase2, ortix2008coulomb}, 2D in \cite{ortix2006frustrated}, and both in \cite{ortix2007screening, ortix2009universality}).

Here, we briefly recap the primary results of these studies for a 2DES and show how the Maxwell construction for phase coexistence is modified. Following Ref.~\cite{ortix2006frustrated}, we assume uniform density within each domain. We comment on the effect of relaxing this assumption below.

One can generically write the free energy per unit area of a mixed phase as
\be 
f=(1-x)f_{1}(n_{1})+xf_{2}(n_{2})+e_{m},
\label{eq:f}
\ee
where $e_{m}$ is the energy density of mixing, which contains both electrostatic and surface energy terms. The energy $e_{m}$ has been considered in detail in Ref.~\cite{ortix2006frustrated} for both the droplet and the stripe geometry. For the case of droplet configurations, $e_m$ is well-approximated by
\be 
e_m = \frac{8}{\sqrt{3}} |n_2 - n_1| \sqrt{\frac{e^2 \sigma}{\epsilon}} x (1-x),
\label{eq: em}
\ee 
where $\sigma$ is the surface tension, and $\epsilon$ is the dielectric constant. (Refer to the Supplemental Material for a brief derivation of Eq.~\ref{eq: em}, as well as an analysis of the stripe geometry contained therein, which yields identical results up to overall numerical coefficients.) The corresponding optimal droplet size $R_d$ is
\be
R_d  = \frac{\sqrt{e^2 \sigma/ \epsilon }}{\left(e^2/\epsilon\right)|n_2 - n_1|}  \frac{\sqrt{3}}{2\sqrt{x\left(1-x\right)}}.
\label{eq:dropletR}
\ee
(A similar expression was presented in the context of quantum Hall stripe and bubble phases in Ref.~\cite{fogler2001stripe}.)

\begin{figure}[htb]
\centering
\includegraphics[scale=0.3]{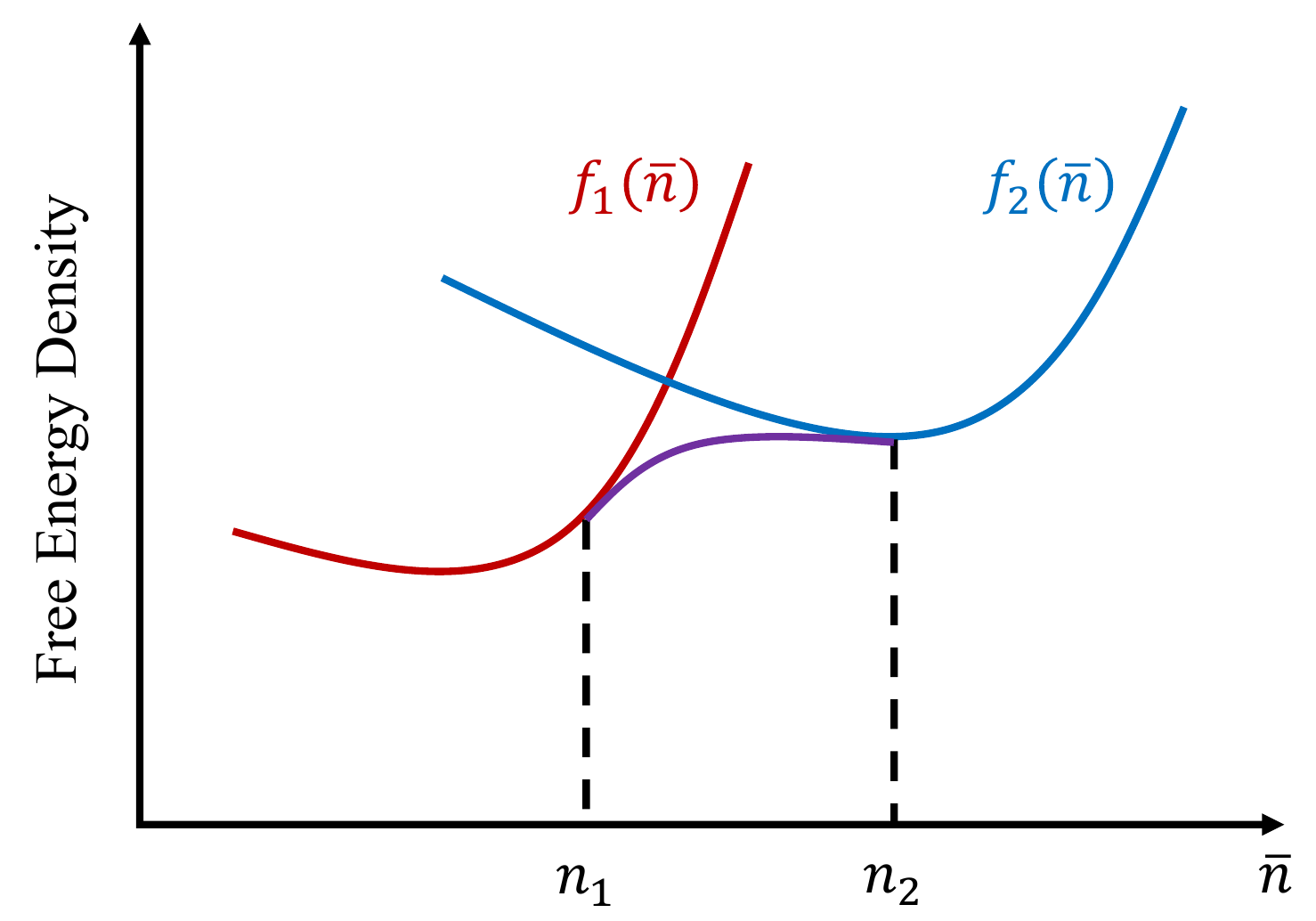}
\caption{Schematic of the modified Maxwell construction accounting for long-range interactions. As in Fig.~\ref{fig:MC}, the red and blue curves represent the free energy per unit area of phases $1$ and $2$, respectively, as a function of the average density $\bar{n}$. The phase coexistence curve (purple line) exhibits a difference in chemical potential between the two phases, as given by Eq.~\ref{eq:MME1}. The corresponding window $\Delta n = n_2 - n_1$ of phase coexistence is smaller than in the conventional Maxwell construction (Fig.~\ref{fig:MC}).
}
\label{fig:MMC}
\end{figure}

Thus, both the droplet size and the energy of mixing depend on the parameter $\beta \equiv \sqrt{e^2 \sigma / \epsilon}$, which has units of energy. One can think of $\beta$ as a parameterization of surface tension: large $\beta$ favors large domains so that the total amount of surface area in the sample is smaller, while small $\beta$ favors small domains. Minimizing the free energy in Eq.~\ref{eq:f} with respect to ${\delta n= n_2-n_1}$ (as in the Maxwell construction)\footnote{Similarly, minimizing the free energy in Eq.~\ref{eq:f} with respect to $x$ produces a relation between the difference in pressure between the two phases \cite{ortix2006frustrated}.}, we arrive at the following relation:
\be
\mu_{1}(n_{1})-\mu_{2}(n_{2})=\frac{8}{\sqrt{3}}\beta.
\label{eq:MME1}
\ee
That is, while in the conventional Maxwell construction, the two phases have the same chemical potential, the inclusion of the term $e_m$ implies that two coexisting phases have a difference in chemical potential $\sim \beta$.\footnote{ We also note that minimizing Eq.~\ref{eq:f} gives densities $n_{1}$ and $n_2$ that vary as a function of the area fraction $x$, unlike the conventional Maxwell construction, in which the densities of the two phases remain fixed as the average density is varied.} This difference in chemical potential can be interpreted as a result of electrochemical equilibrium, in which the difference in electrostatic potential between neighboring domains must be compensated by a difference in the intrinsic chemical potentials. The modified Maxwell construction is depicted schematically in Fig.~\ref{fig:MMC}.

\smallskip

\textit{\textbf{Upper bound on $\Delta n$.}}
Having outlined the modified Maxwell construction, we now consider the bound it places on the window of phase coexistence. The previous analysis provides two fundamental criteria for the presence of a microemulsion phase over any finite interval $\Delta n$ of the average density $\bar{n}$. First, there is a maximum value for the difference in chemical potential between the two phases, effectively determining an upper limit for $\beta$. In a scenario for which phase coexistence is barely possible (i.e., it exists over a very small window of $\bar{n}$), the two phases have densities that are very close to the value $n_c$ at which the two phases have the same energy, $f_1(n_c) = f_2(n_c)$. In this scenario, the difference in chemical potential is near its maximum value, $\Delta \mu=\mu_1(n_c) - \mu_2(n_c)$. Correspondingly, phase coexistence is only permitted when
\be 
\beta < \frac{\sqrt{3}}{8}\Delta \mu  .
\label{eq:coexist1}
\ee 
Equation \ref{eq:coexist1} is derived under the assumption that each domain has a spatially uniform density. In fact, if $\beta$ is large compared to $\Delta \mu$, then this uniform density approximation breaks down \cite{ortix2007screening, jamei2005universal}, and one should instead do a more careful calculation that allows for the charge density within each domain to increase near the domain walls. For large $\beta/\Delta \mu$, this calculation can produce a stable microemulsion phase, but only over a window of density that is exponentially small in the parameter $\beta/\Delta \mu$ \cite{ortix2007screening}. We can thus conclude that in order for there to be a significant window of phase coexistence, $\beta$ must be small enough that Eq.~\ref{eq:coexist1} must not be strongly violated.

On the other hand, if $\beta$ is too small, then the length scale associated with microemulsion droplets becomes unrealistically small. Specifically, phase coexistence is not possible unless the corresponding minimal droplet size (the minimum value of $R_d$ is at $x=1/2$ in Eq.~\ref{eq:dropletR}) is much larger than the inter-electron spacing. That is, $ R_d \gg  n_{1,2}^{-1/2} \approx n_c^{-1/2}$. This condition produces strong inequality
\be
n_{c}^{-1/2}\ll  \frac{\sqrt{3}\, \beta}{\left(e^2/\epsilon\right)\left| n_2 - n_1\right|}.
\label{eq:bnd1}
\ee
Equation \ref{eq:bnd1} sets a \emph{lower} bound on the parameter $\beta$:
\be 
\frac{e^{2}|n_{2}-n_{1}|}{\sqrt{3}\epsilon n_{c}^{1/2}} \ll \beta.
\label{eq:coexist2}
\ee

Combining Eq.~\ref{eq:coexist1} with Eq.~\ref{eq:coexist2} produces an upper bound for the difference in density between the two phases:
\be 
\frac{\Delta n}{n_{c}} \ll \frac{3}{8}\frac{\Delta\mu}{e^{2}n_{c}^{1/2}\big/\epsilon},
\label{eq:deltan}
\ee
Equation \ref{eq:deltan} constitutes the main result of this note. We note that this result is generic in the sense that it does not make any assumptions regarding the nature of the electronic phases involved. Below, we comment on two specific examples: the transition from a Wigner crystal to a Fermi Liquid at zero magnetic field and the phase transition between different bubble phases in high Landau levels.

We emphasize that Eq.~\ref{eq:deltan} makes no assumptions about the value of the surface tension parameter $\beta$. In a given system, the size $\Delta n$ of the microemulsion window, of course, depends on $\beta$, but even without knowing the value of $\beta$, one can place a strong upper bound on $\Delta n$ by combining the requirements that the system should reach electrochemical equilibrium and that droplets should be larger than the interelectron spacing. Notice also from Eq.~\ref{eq:deltan} that, for a fixed chemical potential difference $\Delta \mu$ between the two phases, increasing the dielectric constant $\epsilon$ leads to a wider window of phase coexistence since it reduces the associated Coulomb energy cost. 

\smallskip

\textit{\textbf{Wigner crystal to Fermi liquid transition.}} We now use the inequality in Eq.~\ref{eq:deltan} to make estimates for the range of phase coexistence in the well-studied case of the quantum melting of the Wigner crystal (WC) phase. For this estimate, we use the numerically fitted formulas presented in Ref.~\cite{drummond2009phase}, derived from quantum Monte Carlo methods, for the energy per electron of the FL and WC phases. The resulting estimated upper bounds on $\Delta n / n_c$ are given in Table~\ref{table: WCFL}. Ref.~\cite{drummond2009phase} considers WC trial states with both ferromagnetic (FM) and antiferromagnetic (AFM) spin ordering and suggests a narrow window of the interaction parameter $r_s$ for which the AFM WC has the lowest energy. For completeness, we provide estimates of $\Delta n/n_c$ for all three possible transitions. A similar analysis is presented for the case of phase transitions between bubble phases in the Landau levels of graphene in the Supplemental Material. 
In that case, we find that microemulsion phases occupy a range of filling factors $\delta \nu$ of order $10^{-2}$. The transition between such bubble phases has recently been explored experimentally by Yang et.~al.~\cite{yang2023cascade}.

\begin{table}[htb]
\centering
 \begin{tabular}{|c | c | c |} 
 \hline
Transition & Critical $r_s$ & $\max \Delta n/n_c$  \\ [0.5ex] 
 \hline
  PM FL to FM WC  & $32.89$ & $1.16\times10^{-3}$ \\ [1ex] 
 \hline
  PM FL to AF WC  & $31.26$ & $8.45\times10^{-4}$ \\ 
  \hline
  AF WC to FM WC  & $37.40$ & $3.04\times10^{-4}$ \\
  \hline
 \end{tabular}
 \caption{The bounds on $\Delta n/n_c$ for various WC-FL transitions are tabulated using Eq.~\ref{eq:deltan} and numerical results for the chemical potential of each phase from \cite{drummond2009phase}. Here, FM, AF, and PM represent the Ferromagnetic, Antiferromagnetic, and Paramagnetic phases, respectively.}
\label{table: WCFL}
\end{table}

In Table II we estimate the value of $\Delta n$, in absolute units, for different 2DES systems for which a WC-FL transition has been reported (see Table.~\ref{table: parameters}). In each case, $\Delta n$ is of order $10^7$ cm$^{-2}$, which is generally too small to be observed in current experiments.

\begin{table}[htb]
\centering
 \begin{tabular}{|c| c| c| c|  } 
 \hline
System & $n_c$ (cm$^{-2}$) & Ref &  $\Delta n$  (cm$^{-2}$)\\ [0.5ex] 
 \hline
  MgZnO/ZnO &  $1.6\times10^{10}$ &\cite{falson2022competing}& $1.85\times10^7$\\ 
  \hline
  AlAs & $2\times10^{10}$ & \cite{hossain2020observation}&$2.32\times10^7$\\
  \hline
  GaAs (holes) & $0.77\times10^{10}$& \cite{yoon1999wigner}& $0.89\times10^7$  \\ [1ex] 
 \hline
 \end{tabular}
 \caption{The maximum density window of microemulsion phase coexistence between WC and FL phases is estimated for different experimental 2DES platforms. The listed value of $\Delta n$ corresponds to the PM FL to FM WC transition (the first row in Table.~\ref{table: WCFL}).}
\label{table: parameters}
\end{table}

\smallskip

\textit{\textbf{Disorder-induced density fluctuation.}}
So far, we have been considering a non-disordered 2DES, and we have shown that the maximum value of $\Delta n$ for such systems is very small, on the order of $10^7$ cm$^{-2}$ for the WC-FL transition. In real experiments, the disorder tends to produce a much larger modulation of the density, even in the cleanest samples. The finite width of the WC-FL transition seen in experiments (e.g., Refs.~\cite{falson2022competing, hossain2020observation, yoon1999wigner, knighton2018evidence}) is, therefore, more likely to be attributable to disorder rather than to microemulsion physics. 

Specifically, disorder creates a spatially varying electric potential so that in equilibrium, the local chemical potential also varies spatially. When the electron density is close to $n_c$, the resulting modulation in chemical potential produces puddles of the minority phase embedded in the majority phase \cite{shklovskii_completely_1972, ando1982electronic, shklovskii2007simple, adam_self-consistent_2007}, which appears experimentally as an effective smearing of the transition. For example, a common source of disorder in experimental samples is a three-dimensional concentration $N_\text{imp}$ of impurity charges embedded in the substrate, which produces relatively large and long-wavelength fluctuations of the 2DES density even when $N_\text{imp}$ is low \cite{skinner2013theory}. One can estimate the corresponding value of $\Delta n$ using a self-consistent Thomas-Fermi approximation \cite{ando1982electronic}, which gives $\Delta n_\text{imp} \sim (|dn/d\mu|\,e^{2}N_\text{imp}\big/\epsilon)^{1/2}$ \cite{skinner2013theory}. In the case of a 2DES at large $r_s$, $|dn/d\mu|$ is of the order of $\epsilon\,n_c^{1/2}/e^2$, resulting in $\Delta n_\text{imp} \sim (N_{imp}^2 n_c)^{1/4}$. In the case of MgZnO/ZnO, for example, $\Delta n_\text{imp}$ was estimated to be on the order of several times $10^{9}$ cm$^{-2}$ even for a bulk impurity concentration as low as $10^{14}$ cm$^{-3}$ ($\sim 10$ parts per billion) \cite{falson2022competing}.

\smallskip

\textit{\textbf{Closing remarks.}}
We have shown how one can arrive at an upper bound for the window $\Delta n$ of phase coexistence by combining two criteria: (i) the existence of an electrochemical equilibrium between the two phases, which puts an upper bound on the surface tension parameter $\beta$; and (ii) the requirement that the size of a domain be larger than the interelectron spacing, which puts a lower bound on $\beta$. Combining these two inequalities produces Eq.~\ref{eq:deltan}, which is a seemingly generic upper bound on $\Delta n$ that depends only on the difference in chemical potential $\Delta \mu$ between the two phases at the critical point. (An equivalent expression for the context of quantum Hall phase transitions is presented in the Supplemental Material.) For the case of the WC-FL transition, we use numerical results for $\Delta \mu$ \cite{drummond2009phase} to arrive at an upper bound $\Delta n / n_c \lesssim 10^{-3}$. This very small window of density is consistent with quantum Monte Carlo studies, which have failed to find evidence for a regime of phase coexistence at the WC-FL transition \cite{drummond2009phase, clark2009hexatic}. 

We emphasize that the inequality we derive only applies to phase coexistence in the sense of mesoscopic domains. That is, one can only talk about ``phase 1 coexisting with phase 2'' if the domains of each phase are much larger than the interelectron spacing so that, locally, each region resembles the corresponding bulk phase. However, we cannot rule out the possibility that an instability toward microscopic level mixing (a violation of Eq.~\ref{eq:bnd1}) implies the existence of a new phase in the vicinity of $n_c$, distinct from either of the bulk phases 1 or 2. (For example, one can think of the quantum Hall stripe and bubble phases \cite{Koulakov1996charge, fogler1996ground, fogler2001stripe, fogler1997laughlin} as similar to a ``microemulsion" phase with modulation of density at the scale of the interelectron spacing.) Such new phases cannot be described by a Maxwell construction-type argument.

Let us now comment on the effect of allowing the distance $d$ between the plane of the 2DES and the gate electrode to be finite.  So far, we have adopted the usual assumption that the electron charge is neutralized by a uniform, coplanar sheet of positive background charge. This assumption is valid when $d$ is much larger than the size of the microemulsion domains so that any modulation in the surface charge density of the gate electrode is very weak. In general, however, when $d$ is finite, the long-ranged part of the Coulomb interaction is truncated due to the formation of image charges in the gate electrode, such that at distances $r \gg d$, the electron-electron interaction becomes an effectively short-ranged dipole-dipole interaction, $V(r) \simeq 2 e^2 d^2/r^3$. Thus, a microemulsion domain of size $R_d \gg d$ has a finite electric potential in its interior even in the limit $R_d \rightarrow \infty$ (unlike in the case of the unscreened Coulomb potential, for which the electric potential diverges, as mentioned above). Requiring equilibration of the electrochemical potential between the interior of two such large domains gives \cite{spivak2003phase, spivak2004phases} $\Delta \mu = (e^2 d \Delta n)/\epsilon$, where the right-hand side of this expression represents the difference in electric potential energy between the two domains. Consequently, the difference in density $\Delta n$ between the two phases can be no smaller than $(\Delta \mu)/(e^2 d/\epsilon)$.

Notice, however, that this lower bound for $\Delta n$ is much smaller than the bound given by Eq.~\ref{eq:deltan} unless $d \lesssim n_c^{-1/2}$. For such small $d$, even the interaction between nearest-neighboring electrons is of a dipole type, and the bulk phases are strongly modified. For the case of the WC-FL transition, for example, the WC phase is completely eliminated by the gate screening when $d \ll n_c^{-1/2}$ (where $n_c$ denotes the critical density without gate screening) \cite{spivak2006transport, skinner2010anomalously, skinner2010simple}. Thus, the upper bound given by Eq.~\ref{eq:deltan} still applies even when $d$ is allowed to be finite.

\ 

\textit{\textbf{Acknowledgments:}}
The authors are grateful to Steve Kivelson, Boris Spivak, and Boris Shklovskii for helpful discussions and to Andrea Young and Fangyuan Yang for a related collaboration. This work was supported by the NSF under Grant No.~DMR-2045742. 
\bibliography{micro}

\clearpage
\widetext
\begin{center}
\textbf{\large Supplementary Information for ``Upper bound on the window of density occupied by microemulsion phases in 2D electron systems"}
\end{center}
\setcounter{equation}{0}
\setcounter{figure}{0}
\setcounter{table}{0}
\setcounter{page}{1}
\makeatletter
\renewcommand{\theequation}{S\arabic{equation}}
\renewcommand{\thefigure}{S\arabic{figure}}
\renewcommand{\bibnumfmt}[1]{[S#1]}
\renewcommand{\citenumfont}[1]{S#1}

\begin{center}
Sandeep Joy, Brian Skinner \\
\textit{Department of Physics, Ohio State University, Columbus, OH 43210, USA} \\
(Dated: \today)
\end{center}
\section{S1. Electrostatic energy of a microemulsion droplet}
\label{sec: ele}

\begin{figure}[htb]
\centering
\includegraphics[scale=0.45]{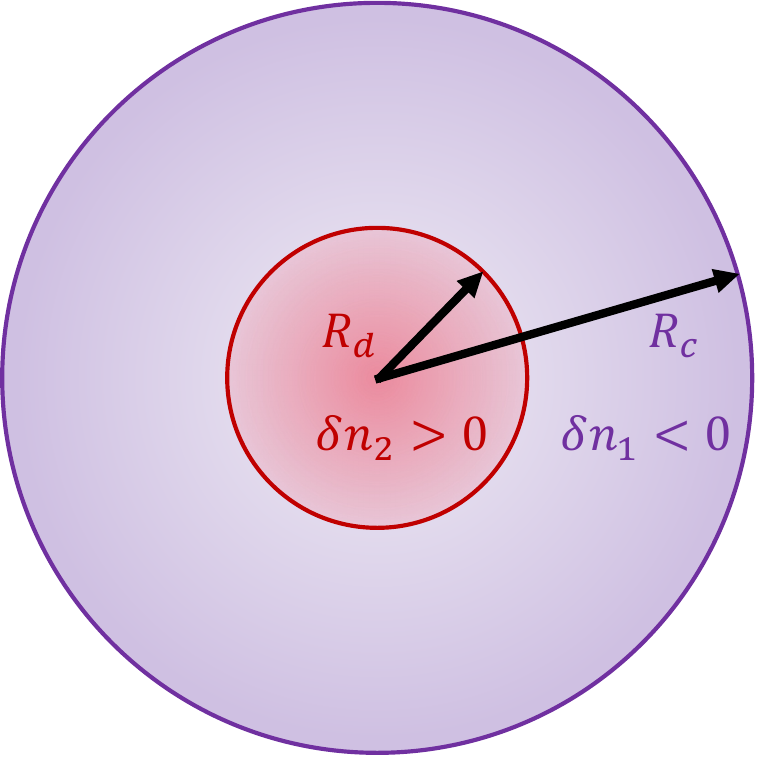}
\caption{Diagram representing the construction of Wigner-Seitz cell. The background density $-\bar{n}<0$ ensures the charge neutrality of the entire cell. Inside the inner disk, the total density is given by $\delta n_{2}>0$. In the outer shell, the total density is given by $\delta n_{1}<0$.}
\label{fig:WZcell}
\end{figure}

Here, we provide a calculation of the mixing energy for a microemulsion phase in which the minority phase (which we take to be phase 2) forms droplets embedded in the majority phase (phase 1). To model this system, we use a Wigner-Seitz cell approximation (see Fig.~\ref{fig:WZcell}), dividing the entire sample into approximately circular cells of radius $R_c$. The total number of cells is $N_{c}=A/\pi R_c^{2}$, where $A$ is the total area of the sample and $R_d$ is the radius of the inner disk. The area fraction of phase 2 is represented by $x=(R_d/R_c)^{2}$. The droplet geometry is applicable when $x$ is close to either $0$ or $1$. When $x$ is close to $1/2$, a stripe geometry has lower energy. Nevertheless, one can use the droplet calculation to arrive at an approximate formula that interpolates across the entire range of $x$ \cite{ortix2006frustrated}, as we discuss further below. 

To ensure that each cell is charge neutral, we introduce a uniformly charged background with a density of $-\bar{n}$ (where $\bar{n}=(1-x)n_1+xn_2$). Within this model, the local density is given by:
\begin{equation}
\delta n(\rho)=\left\{ \begin{array}{cc}
\delta n_{2} & 0<\rho<R_{d}\\
\delta n_{1} & R_{d}<\rho<R_{c},
\end{array}\right.    
\end{equation}
where $\rho$ is the radial coordinate in the plane, $\delta n_{2}\equiv n_{2}-\bar{n}$, and $\delta n_{1}=n_{1}-\bar{n}$. 

We evaluate the electrostatic energy using the electrostatic potential $V(\rho)$ of a uniform charge disk of radius $R$ and charge density $\sigma$ \cite{ciftja2011the}:
\be
V(\rho)=\left\{ \begin{array}{cc}
\frac{4\sigma}{\epsilon}E\left[\left(\frac{\rho}{R}\right)^{2}\right], & 0<\rho<R\\
\frac{4\sigma}{\epsilon}\left(\left(\frac{\rho}{R}\right)E\left[\left(\frac{\rho}{R}\right)^{2}\right]+\frac{\left(1-\left(\rho/R\right)^{2}\right)}{\left(\rho/R\right)}K\left[\left(\frac{\rho}{R}\right)^{2}\right]\right), & R<\rho<\infty
\end{array}\right.
\ee
The electrostatic energy of a droplet is then given by:
\be
E_c = \frac{8\pi e^2(n_2 - n_1)^2 R_c^3}{3\epsilon} f(x)
\ee
where $f(x)$ is the function
\be
f(x) = x\left(\sqrt{x}+x-(1+x)E(x)+(1-x)K(x)\right).
\label{eq:f(x)}
\ee
Here, $K(x)$ and $E(x)$ are the complete elliptic integrals of the first and second kind, respectively. 

As mentioned above, the configuration in which drops of phase 2 are embedded in phase 1 appears only when $x$ is sufficiently close to zero. At small $x$, Eq.~\ref{eq:f(x)} gives $f(x) \simeq x^{3/2}$. At $x$ close to $1$, the roles of phases 1 and 2 are reversed, implying $f(x) \simeq (1-x)^{3/2}$. Following the results of Ref.~\cite{ortix2006frustrated}, these two limits imply an approximate expression $f(x) \approx \left( x (1-x) \right)^{3/2}$ for the entire range of $x$. 

By minimizing the sum of the total Coulomb energy ($N_c\times E_c$) and the total surface energy ($N_c\times 2\pi\sigma R_d$) with respect to $R_d$, we obtain the result given in Eq.~8 and Eq.~9 from the main text.

\section{S2. Coulomb-frustrated phase coexistence in the stripe geometry}
\label{sec: stripes}
In this Appendix, we re-derive the bound on $\Delta n$ assuming a stripe configuration rather than the droplet geometry that was assumed in the main text. In the stripe geometry, the energy density of mixing $e_m$ and corresponding optimal stripe width $R_s$ is well-approximated by \cite{ortix2006frustrated}
\be 
e_m = |n_2 - n_1| \sqrt{\frac{e^2 \sigma}{\epsilon}} 2\sqrt{2 }x (1-x)\left(-\log x (1-x)\right)^{1/2},
\label{eq: stripes_em}
\ee 
and
\be
R_s  = \frac{\sqrt{e^2 \sigma/ \epsilon }}{\left(e^2/\epsilon\right)|n_2 - n_1|}  \frac{1}{\sqrt{2}x (1-x)\left(-\log x (1-x)\right)^{1/2}}.
\label{eq: stripes_R}
\ee
The associated modified Maxwell equation for the stripe geometry is given by
\be
\mu_{1}(n_{1})-\mu_{2}(n_{2})= 2\sqrt{2}\left(-\log x (1-x)\right)^{1/2}\beta.
\label{eq: stripes_MME1}
\ee
In contrast to the scenario involving droplets, for the stripe geometry, the modified Maxwell equation explicitly depends on $x$. It must be noted that the right-hand side of Eq.~\ref{eq: stripes_MME1} exhibits divergence as $x$ approaches zero (or one). However, these divergences are not relevant physically because, at such small values of $x$ (or $1-x$), the droplet geometry has a lower energy $e_m$.

Following similar reasoning as presented in the main text, it becomes evident that there exists a maximum value for $\beta$ that allows for phase coexistence:
\be 
\beta < \frac{\Delta \mu}{ 2\sqrt{2}\left(-\log x (1-x)\right)^{1/2}}  .
\label{eq: stripes_coexist1}
\ee 
On the other hand, phase coexistence is not possible unless the corresponding optimal size of the stripe is much larger than the inter-electron spacing. That is, $ R_s \gg  n_{1,2}^{-1/2} \approx n_c^{-1/2}$, suggesting
\be
n_{c}^{-1/2}\ll   \frac{\beta}{\left(e^2/\epsilon\right)|n_2 - n_1|}  \frac{1}{\sqrt{2}x (1-x)\left(-\log x (1-x)\right)^{1/2}}.
\label{eq: stripes_bnd1}
\ee
Rearranging the above equation, we arrive at
\be
\frac{e^2 |n_2 - n_1|}{\epsilon n_c^{1/2}}\sqrt{2}x (1-x)\left(-\log x (1-x)\right)^{1/2} \ll \beta.
\label{eq: stripes_coexist2}
\ee
Combining Eq.~\ref{eq: stripes_coexist1} and Eq.\ref{eq: stripes_coexist2}, we arrive at the following result
\be
\frac{\Delta n}{n_c} \ll \frac{\Delta \mu}{(e^{2}n_{c}^{1/2}\big/\epsilon)} \frac{1}{4 x (1-x)\left(-\log x (1-x)\right)} .
\ee
The inequality on the right-hand side is most stringent when $x=1/2$, for which we obtain the result
\be
\frac{\Delta n}{n_c} \ll \frac{1}{2\log 2}\frac{\Delta \mu}{(e^{2}n_{c}^{1/2}\big/\epsilon)}  .
\ee
Thus, up to an overall numerical factor, we replicate the primary result presented in the main text (see Eq.~14).

\section{S3. Density range of microemulsion for bubble phases: Illustration using the $n=3$ Landau level in graphene}
\label{sec: grapheneLL}

Here, as an example of the generality of our result, we use Eq.~14 (from the main text) to place a bound on the range of filling factors associated with the phase coexistence between the so-called bubble phase in the $n = 3$ Landau Level \cite{Koulakov1996charge, fogler1996ground}. Inserting units of the magnetic length, we arrive at an alternate formulation of Eq.~14 (from the main text):
\be 
\Delta \nu \ll \frac{3 \sqrt{2 \pi}}{8} \frac{ \Delta \mu(\nu_c)}{e^2/\epsilon l_B} \nu_c^{1/2}.
\label{eq:deltanu}
\ee 
In this equation, $\Delta \nu$ denotes the interval of filling factors corresponding to phase coexistence, $\nu_c$ signifies the partial filling factor at which the energies of the two phases are equal, and $\Delta \mu(\nu_c)$ represents the discrepancy in chemical potential between the phases at $\nu = \nu_c$. As an example, the results for $n=3$ Landau levels in graphene are presented in Table~\ref{table: Graphene3nLL}, with results for the energy per electron taken from Ref.~\cite{goerbig2004competition}.

\begin{table}[htb]
\centering
 \begin{tabular}{|c | c | c |} 
 \hline
Transition & $\nu_c$ & $\max \Delta \nu/\nu_c$ (Eq.~\ref{eq:deltanu})  \\ [0.5ex] 
 \hline
  $M=1$ to $M=2$   & $0.18$ & $2.25\times10^{-2}$ \\ 
  \hline
  $M=2$ to $M=3$  & $0.28$ & $2.56\times10^{-2}$ \\
 \hline
 \end{tabular}
 \caption{The bounds on $\Delta \nu/\nu_c$ for the phase transition between bubble phases in the $n=3$ Landau levels of graphene \cite{goerbig2004competition}. The integer $M$ labels the number of electrons per bubble. }
\label{table: Graphene3nLL}
\end{table}

\end{document}